\documentclass{camera}
\usepackage{graphicx}

\begin{document}

\title{Modeling the host galaxies of long-duration gamma-ray bursts}

\author{Emily M. Levesque$^1$}

\organization{Institute for Astronomy, University of Hawaii, 2680 Woodlawn Dr., Honolulu, HI 96822}

\maketitle

\begin{abstract}
We present the first results of our investigation into the ISM environments of long-duration GRB (LGRB) host galaxies. We apply a new suite of stellar population synthesis and photoionization models to new, uniform, rest-frame optical observations of eight LGRB host galaxies ranging from $z = 0.01$ to $z = 0.81$. We also compare these hosts to a variety of local and intermediate-redshift galaxy populations. We find that LGRB host galaxies generally have low-metallicity ISM environments. As a whole, the ISM properties of our LGRB hosts set them apart from the general galaxy population, host galaxies of nearby Type Ic supernovae, and nearby metal-poor galaxies. With these comparisons we investigate whether LGRB host galaxies may be used as accurate tracers of star formation in distant galaxies.
\end{abstract}

\footnotetext[1]{Predoctoral Fellow, Smithsonian Astrophysical Observatory}
\section{Introduction}
Long-duration gamma-ray bursts (LRGBs) are often proposed as tools for tracing star-formation in the high-redshift universe [1,2,3,4,5,6]. These phenomena, with a burst duration of $>$ 2 s, are associated with the core-collapse of rapidly-rotating massive stars and the production of broad-lined Type Ic supernovae [7,8,9,10,11]. However, in recent years a connection between some LGRBs and low-metallicity galaxies has been proposed, which could threaten their utility as unbiased tracers of star formation in the universe [12,13,14,15,16].

It possible that the observed connection between LGRBs and low-metallicity environments may not be a direct result of metallicity at all, but rather an artifact of other galactic properties that might be favored by LGRBs, such as age or burst-like star formation histories [17,18]. To determine whether metallicity was a key environmental factor in LGRB host galaxies, my co-authors and I recently conducted the first uniform high-S/N spectroscopic survey of LGRB host galaxies [19].

\section{Observations and Analyses} Using the Low Resolution Imaging Spectrograph (LRIS) and Near-Infrared Spectrograph (NIRSPEC) on the Keck telescopes at Mauna Kea Observatory, as well as the Low Dispersion Survey Spectrograph 3 on the Magellan Clay telescope at Las Campanas Observatory, we obtained spectroscopic data for 8 $z < 1$ LGRB host galaxies. Combined with two existing spectra of LGRB hosts from the literature, we used these spectra to determine a number of ISM properties for these galaxies using diagnostic ratios of optical emission lines. These parameters included metallicity, ionization parameter, $E(B-V)$, young stellar population age, and star formation rate; a detailed discussion of the diagnostics used to calculate these parameters is given in [19].

\section{Comparison Samples}
After determining ISM properties from our LGRB host galaxy sample, we wished to compare these galaxies to a variety of other star-forming galaxy samples. We included a number of star-forming galaxies at $z < 0.3$ for comparison to our nearby LGRB hosts. These samples included 60,920 star-forming galaxies from Data Release 4 of the Sloan Digital Sky Survey [20,21], 95 star-forming galaxies from the Nearby Field Galaxy Survey [22,23], 36 blue compact galaxies (BCGs) [24], 10 metal-poor galaxies (MPGs) [25], and a sample of 8 broad-lined Type Ic SN host galaxies [15]. We also compared the higher-redshift LGRB host galaxies in our sample to a sample of $0.3 < z < 1$ star-forming galaxies from the Team Keck RedShift Survey [26]. These comparisons were done based on both the derived ISM properties for these samples, such as metallicity (Figure 1), and direct measurements of emission line diagnostic ratios from these sample spectra - for more discussion, see [19]. Finally, we compared the LGRB host spectra to a new grid of stellar population synthesis and photoionization models [27].  The LGRBs showed generally good agreement with the lower-metallicity regime of the models; however, several shortcomings in the model grid were made apparent and must be addressed in future work.

\section{Results and Conclusions}
Based on the comparison of the LGRB host galaxies to samples from the general star-forming galaxy population, we found that LGRB host galaxies have lower-metallicity ISM environments than the general galaxy population, both nearby ($z < 0.3$) and at intermediate redshifts ($0.3 < z < 1$). As a whole, the ISM properties of our LGRB hosts set them apart from the general galaxy population, including the host galaxies of nearby Type Ic supernovae.This suggests that LGRBs may be biased towards low-metallicity environments, and that this difference might be key in distinguishing between the properties of regular Type Ic supernovae progenitors and the enigmatic massive stars that produce LGRBs.


\begin{thebibliography}{00}
\bibitem{1} Wijers, R. A. M. J., Bloom, J. S., Bagla, J. S., \& Natarajan, P. 1998, MNRAS, 297, L13
\vspace{-7pt}
\bibitem{2} Fynbo, J. P. U., Hjorth, J., Malesani, D., Sollerman, J., Watson, D., Jakobsson, P., Gorosabel, J., \& Jaunsen, A. O. 2007, arXiv:astro-ph/0703458v2
\vspace{-7pt}
\bibitem{3} Chen, H-W., Prochaska, J. X., Bloom, J. S., \& Thompson, I. B. ApJ, 634, L25
\vspace{-7pt}
\bibitem{4} Berger, E., Penprase, B. E., Cenko, S. B., Kulkarni, S. R., Fox, D. B., Steidel, C. C., Reddy, N. A. 2006, ApJ, 642, 979
\vspace{-7pt}
\bibitem{5} Prochaska, J. X., Chen, H-W., Dessauges-Zavadsky, M., Bloom, J. S. 2007, ApJ, 666, 267
\vspace{-7pt}
\bibitem{6} Savaglio, S., Glazebrook, K., \& Le Borgne, D. 2009, ApJ, 1091, 182
\vspace{-7pt}
\bibitem{7} Woosley, S. E. 1993, ApJ, 405, 273
\vspace{-7pt}
\bibitem{8} Galama, T. J. et al. 1998, Nature, 395, 670
\vspace{-7pt}
\bibitem{9} Stanek, K. Z. et al.\ 2003, ApJ, 591, L17
\vspace{-7pt}
\bibitem{10} Kawabata, K. S. et al.\ 2003, ApJ, 593, L19
\vspace{-7pt}
\bibitem{11} Hjorth, J. et al. 2003, Nature, 423, 847
\vspace{-7pt}
\bibitem{12} Stanek, K. Z., et al.\ 2006, Acta Astron. 56, 333
\vspace{-7pt}
\bibitem{13} Fruchter, A. S. et al.\ 2006, Nature, 441, 463
\vspace{-7pt}
\bibitem{14} Kewley, L. J., Brown, W. R., Geller, M. J., Kenyon, S. J., \& Kurtz, M. J. 2007, AJ, 133, 882
\vspace{-7pt}
\bibitem{15} Modjaz, M., Kewley, L. J., Kirshner, R. P., Stanek, K. Z., Challis, P., Garnavich, P. M., Greene, J. E., Kelly, P. L., Prieto, J. L. 2008, AJ, 135, 1136
\vspace{-7pt}
\bibitem{16} Kocevski, D., West, A. A., \& Modjaz, M. 2009, ApJ, 702, 377
\vspace{-7pt}
\bibitem{17} Bloom, J. S., Kulkarni, S. R., \& Djorgovski, S. G. 2002, ApJ, 121, 1111
\vspace{-7pt}
\bibitem{18} Berger, E., Fox, D. B., Kulkarni, S. R., Frail, D. A., \& Djorgovski, S. G. 2007, ApJ, 660, 504
\vspace{-7pt}
\bibitem{19} Levesque, E. M., Berger, E., Kewley, L. J., \& Bagley, M. M. 2009, AJ, in press (arXiv:0907.4988)
\vspace{-7pt}
\bibitem{20} Adelman-McCarthy, J. K. et al.\ 2006, ApJS, 162, 38
\vspace{-7pt}
\bibitem{21} Kewley, L. J., Groves, B., Kauffman, G., \& Heckman, T. 2006, MNRAS, 372, 961
\vspace{-7pt}
\bibitem{22} Jansen, R. A., Fabricant, D., Franx, M., \& Caldwell, N. 2000a, ApJS, 126, 331
\vspace{-7pt}
\bibitem{23} Jansen, R. A., Franx, M., Fabricant, D., \& Caldwell, N. 2000b, ApJS, 126, 271
\vspace{-7pt}
\bibitem{24} Kong, X. \& Cheng, F. Z. 2002, A\&A, 389, 845
\vspace{-7pt}
\bibitem{25} Brown, W., Kewley, L. J., \& Geller, M. J. 2008, AJ, 135, 92
\vspace{-7pt}
\bibitem{26} Kobulnicky, H. A. \& Kewley, L. J. 2004, ApJ, 617, 24
\vspace{-7pt}
\bibitem{27} Levesque, E. M., Kewley, L. J., \& Larson, K. L. 2009, AJ, in press (arXiv:0908.0460)
\end{thebibliography}
\end{document}